\documentclass[aps,prb,twocolumn,superscriptaddress]{revtex4-1}
\usepackage[pdftex]{graphicx}
\usepackage{booktabs}
\usepackage{multirow}
\usepackage[mediumspace,mediumqspace,Grey,squaren]{SIunits}
\usepackage{color}
\usepackage{natbib}
\usepackage{amsmath,amssymb}
\usepackage{lipsum}
\usepackage[normalem]{ulem}
\usepackage{bm}
\usepackage[euler]{textgreek}
\graphicspath{ {./Figures/} }

\begin{document} 
\title{Tracking ultrafast photocurrents in the Weyl semimetal TaAs using THz emission spectroscopy}

\author{N. Sirica}
\email{nsirica@lanl.gov}
\affiliation{Center for Integrated Nanotechnologies, Los Alamos National Laboratory, Los Alamos, NM 87545, USA}
\author{R.I. Tobey}
\affiliation{Center for Integrated Nanotechnologies, Los Alamos National Laboratory, Los Alamos, NM 87545, USA}
\affiliation{Zernike Institute for Advanced Materials, University of Groningen, Groningen, The Netherlands}
\author{L.X. Zhao}
\affiliation{Institute of Physics, Chinese Academy of Sciences, Beijing 100190, China}
\author{G.F. Chen}
\affiliation{Institute of Physics, Chinese Academy of Sciences, Beijing 100190, China}
\author{B. Xu}
\affiliation{Institute of Physics, Chinese Academy of Sciences, Beijing 100190, China}
\author{R. Yang}
\affiliation{Institute of Physics, Chinese Academy of Sciences, Beijing 100190, China}
\author{B. Shen}
\affiliation{Department of Physics and Astronomy, University of California, Los Angeles, CA 90095, USA}
\author{D.A. Yarotski}
\affiliation{Center for Integrated Nanotechnologies, Los Alamos National Laboratory, Los Alamos, NM 87545, USA}
\author{P. Bowlan}
\affiliation{Center for Integrated Nanotechnologies, Los Alamos National Laboratory, Los Alamos, NM 87545, USA}
\author{S.A. Trugman}
\affiliation{Center for Integrated Nanotechnologies, Los Alamos National Laboratory, Los Alamos, NM 87545, USA}
\author{J.-X. Zhu}
\affiliation{Center for Integrated Nanotechnologies, Los Alamos National Laboratory, Los Alamos, NM 87545, USA}
\author{Y.M. Dai}
\affiliation{Center for Integrated Nanotechnologies, Los Alamos National Laboratory, Los Alamos, NM 87545, USA}
\affiliation{School of Physics, Nanjing University, Nanjing 210093, China}
\author{A.K. Azad}
\affiliation{Center for Integrated Nanotechnologies, Los Alamos National Laboratory, Los Alamos, NM 87545, USA}
\author{N. Ni}
\affiliation{Department of Physics and Astronomy, University of California, Los Angeles, CA 90095, USA}
\author{X.G. Qiu}
\affiliation{Institute of Physics, Chinese Academy of Sciences, Beijing 100190, China}
\author{A.J. Taylor}
\affiliation{Center for Integrated Nanotechnologies, Los Alamos National Laboratory, Los Alamos, NM 87545, USA}
\author{R.P. Prasankumar}
\email{rpprasan@lanl.gov}
\affiliation{Center for Integrated Nanotechnologies, Los Alamos National Laboratory, Los Alamos, NM 87545, USA}


\begin{abstract}
We investigate polarization-dependent ultrafast photocurrents in the Weyl semimetal TaAs using terahertz (THz) emission spectroscopy. Our results reveal that highly directional, transient photocurrents are generated along the non-centrosymmetric $c$-axis regardless of incident light polarization, while helicity-dependent photocurrents are excited within the $ab$-plane. This is consistent with earlier static photocurrent experiments, and demonstrates on the basis of both the physical constraints imposed by symmetry and the temporal dynamics intrinsic to current generation and decay that optically induced photocurrents in TaAs are inherent to the underlying crystal symmetry of the transition metal monopnictide family of Weyl semimetals.  
\end{abstract}
\maketitle

\textit{Introduction:} 
 The recent prediction and subsequent discovery of Weyl fermions as emergent quasiparticles in materials possessing strong spin-orbit interaction and broken time-reversal or inversion symmetry has generated a great deal of interest, due to their relevance in fundamental physics and applied technology alike \cite{Armitage2018,Hasan2017,Yan2017,Jia2016,Tokura2017}. A defining characteristic of the electronic structure of these so-called Weyl semimetals (WSM) is the existence of Weyl points, where nondegenerate, linearly dispersing bands found in the bulk of these materials cross \cite{Xu2011,Wan2011,Xu2015,Lv2015,Yang2015}. These points act as monopoles of Berry curvature in momentum space, defining the chiral charge of the Weyl fermion, and are found to be topologically stable even in the absence of any particular symmetry \cite{Xu2011,Wan2011}. This leads to several unique experimental manifestations of Weyl physics, including Fermi arcs, which connect the surface projections of two Weyl points having opposite chirality \cite{Xu2015, Lv2015, Yang2015, Liu2015}, and the chiral anomaly, in which a negative magnetoresistance develops from intervalley pumping between Weyl cones of opposite chirality \cite{Parameswaran2014,Huang2015,Zhang2016}. 

 Linear optical spectroscopy has revealed novel phenomena in WSM, including the theoretically predicted linear scaling of conductivity with frequency and strong Weyl fermion-phonon coupling \cite{Xu2016, Xu2017, Kimura2017}. Further insight into Weyl physics can be gained from nonlinear optics through the effect that Berry curvature introduces on such nonlinear quantities as the shift vector and photocurrent \cite{Morimoto2016,Morimoto2016_02}. Specifically, for non-centrosymmetric WSM, like the transition metal monopnictides (TMMP), the shift vector, which defines a difference in the center of electron charge density within one unit cell following optical excitation \cite{Sipe2000}, develops an additional contribution arising from a change in Berry curvature between the bands participating in the transition \cite{Morimoto2016}. This behavior has been studied in the TMMP, where a giant anisotropic nonlinear response was observed in the optical and near-infrared (IR) range that was 200 times larger than that of standard nonlinear crystals like GaAs \cite{Wu2016,Patankar2018}. In that case, the dominant contribution to the nonlinear response measured along the polar $c$-axis was attributed to a helicity-independent shift current originating from the strong polar character of these materials \cite{Patankar2018,Li2018}. However, polarization-dependent photocurrent measurements made on WSM following mid-IR excitation have suggested a topologically non-trivial contribution to the shift current, revealing a colossal bulk photovoltaic effect that may be linked to divergent Berry curvature near the Weyl nodes \cite{Osterhoudt2017,Ma2018}.

 Helicity-dependent photocurrents measured in topological insulators \cite{McIver2011,Kastl2015} and WSM \cite{Ma2017,Kai2017,Ji2018} have likewise provided insight into the topologically non-trivial behavior of these materials. Here, the direction of these photocurrents can be switched by changing light helicity (i.e., degree of circular polarization), opening up the possibility for all-optical control without an external bias field. In WSM, the contribution of injection currents, or photocurrents resulting from an asymmetric distribution of carriers in momentum space due to the interference of different light polarizations \cite{Sipe2000}, to the circular photogalvanic effect (CPGE) gives rise to a helicity-dependent photocurrent that is claimed to provide a direct experimental measure for the topological charge of Weyl points \cite{Ma2017,deJuan2017,Chan2017,Zhang2018}. Experimentally, the CPGE was demonstrated in static photocurrent measurements of the WSM TaAs following mid-IR and optical excitation \cite{Ma2017, Kai2017}, and was subsequently used to determine Weyl fermion chirality based upon the specific direction that current flows relative to the high symmetry axes of the crystal. Despite their observation of a helicity-dependent photocurrent, ref. \onlinecite{Ma2017} reported a negligible contribution from shift currents, even when measuring along the non-centrosymmetric $c$-axis. This finding contrasts with that in ref. \onlinecite{Osterhoudt2017}, raising the question of why static photocurrent measurements made on the same WSM reveal such different results.  
 
 In this letter, we demonstrate the generation of both helicity-dependent and helicity-independent ultrafast photocurrents as measured by terahertz (THz) emission spectroscopy on the WSM TaAs. THz emission, detected either directly though electro-optic sampling (EOS) or by THz field-induced second harmonic generation (TFISH), is a contact-free means of measuring transient photocurrents on the intrinsic timescales that underlie their generation and decay \cite{THzSpecBook}. Despite our use of femtosecond, near-IR optical pulses to drive these photocurrents, our results agree well with previous static photocurrent measurements, and have the added advantage over such experiments in that photo-thermal effects are largely mitigated due to the ultrashort duration of the driving pulse. Below, we will focus on the results obtained from TaAs, but the same behavior is observed for the closely related TMMP WSM NbAs (Supplementary Fig. 1)\cite{Supp}.  

\begin{figure}[t]
    \centering     
    \includegraphics[width=\columnwidth]{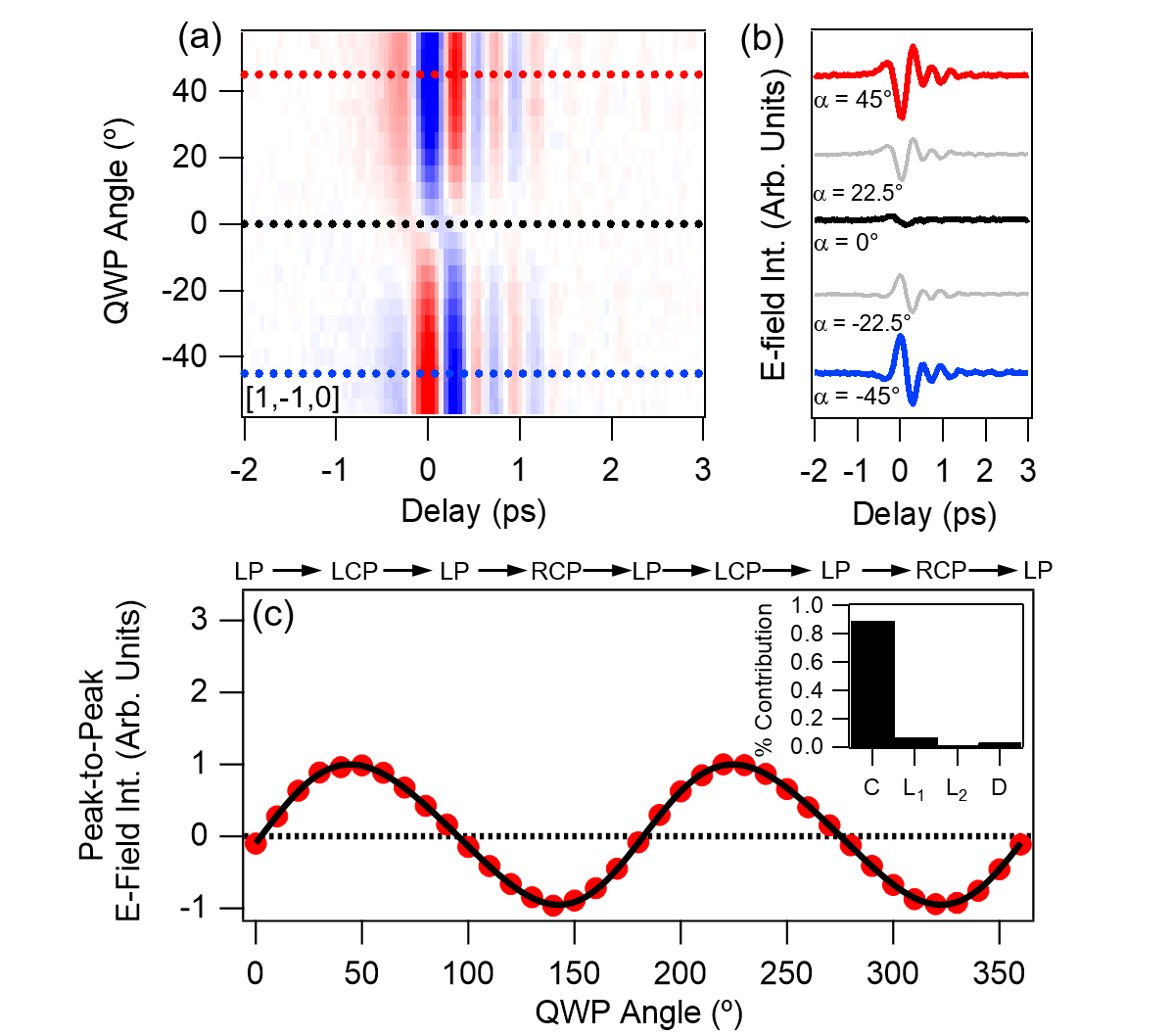}
    \caption{(a) False color plot and (b) select time-dependent THz traces, illustrating the polarity reversal of the emitted THz waveform upon changing the helicity of the optical generation pulse. Traces shown in (b) are obtained using quarter waveplate (QWP) angles of $\pm45^\circ$, $\pm22.5^\circ$, and $0^\circ$, which correspond to right/left circular, elliptical or linear polarizations, respectively. (c) Peak-to-peak E-field amplitude plotted as a function of $\alpha$ and fit with Eq. (1). The inset illustrates the relative weight of each fitting component.}
    \label{fig:CPGE}
\end{figure}

\textit{Experimental:} 
 THz emission from a 1 mm thick as-grown TaAs single crystal was measured using an amplified Ti:sapphire laser system operating at a 1 kHz repetition rate. Ultrashort optical pulses centered at 800 nm (1.55 eV) with a duration of $\sim40$ femtoseconds (fs) and fluences up to 17 mJ/cm$^2$ were incident on the crystal surface, and the specularly emitted THz radiation was detected by free space EOS in a 0.5 mm thick $<110>$ ZnTe crystal (Supplementary Fig. 2) \cite{Supp}. Measurements were made on the (001) and (112) faces at both $\sim5^\circ$ and $\sim45^\circ$ angles of incidence. The (112) surface, which has been the subject of previous investigations into the nonlinear optical properties of these materials \cite{Wu2016,Patankar2018}, possesses two in-plane, high symmetry axes, [1,-1,0] and [1,1,-1] (Supplementary Fig. 2), where the latter contains a projection of the inversion symmetry-broken $c$-axis. A wire grid polarizer was used to determine the polarization of the emitted THz pulses relative to these crystal axes. Finally, all experiments were performed at room temperature in an enclosure purged with dry air.

\textit{Results and Discussion:} 
 Our main results are shown in Fig. \ref{fig:CPGE}, which illustrates in both the false color image (Fig. \ref{fig:CPGE}(a)) and select stacked traces (Fig. \ref{fig:CPGE}(b)) a clear polarity reversal of the emitted THz waveform polarized along the [1,-1,0] axis, occurring as the helicity of the optical generation pulse is tuned from left circular to right circular polarization. Analysis of the THz waveforms in Fig. \ref{fig:CPGE} shows a $180^\circ$ polarity reversal, with no variation in frequency, and a change in amplitude that corresponds to the degree of ellipticity of the incident light pulse. A plot of the peak-to-peak amplitude of the emitted THz electric (\textbf{E})-field while rotating the \textlambda /4 waveplate (QWP) over a full $360^\circ$ reveals a sinusoidal dependence whose periodicity matches a change in helicity of the incident light (Fig. \ref{fig:CPGE}(c)). Fitting with a general expression for the polarization dependence of the photocurrent \cite{McIver2011}, 
 \begin{equation}
     j(\alpha) = C\sin{2\alpha}+L_1\sin{4\alpha}+L_2\cos{4\alpha}+D,
     \label{Equation_1}
 \end{equation}
 where $\alpha$ is the QWP angle, reveals the dominant ($\sim90\%$) contribution to arise from the helicity-dependent term, \textit{C}. However, the emitted THz pulse is strongly suppressed, but not entirely quenched, when the polarization of the incident light is linear (Fig. \ref{fig:CPGE}(b)). This implies a small deviation of $\sim7\%$ from the ideal $\sin{2\alpha}$ behavior, which is due to the helicity-independent, but linearly-dependent term \textit{L\textsubscript{1}}, as well as an $\sim3\%$ contribution from the polarization-independent term \textit{D}. Further investigation into the linearly-dependent THz emission reveals a change in both amplitude and phase of the THz waveform as the polarization of the generating pulse is tuned from horizontal to vertical (Supplementary Fig. 3) \cite{Supp}. However, since both \textit{L\textsubscript{1}} and \textit{D} provide only small contributions to the polarization dependence of the emitted THz pulse along the [1,-1,0] axis, we will primarily focus on the dominant, helicity-dependent behavior observed along this high-symmetry direction.

\begin{figure}[t]
    \centering     
    \includegraphics[width=0.9\columnwidth]{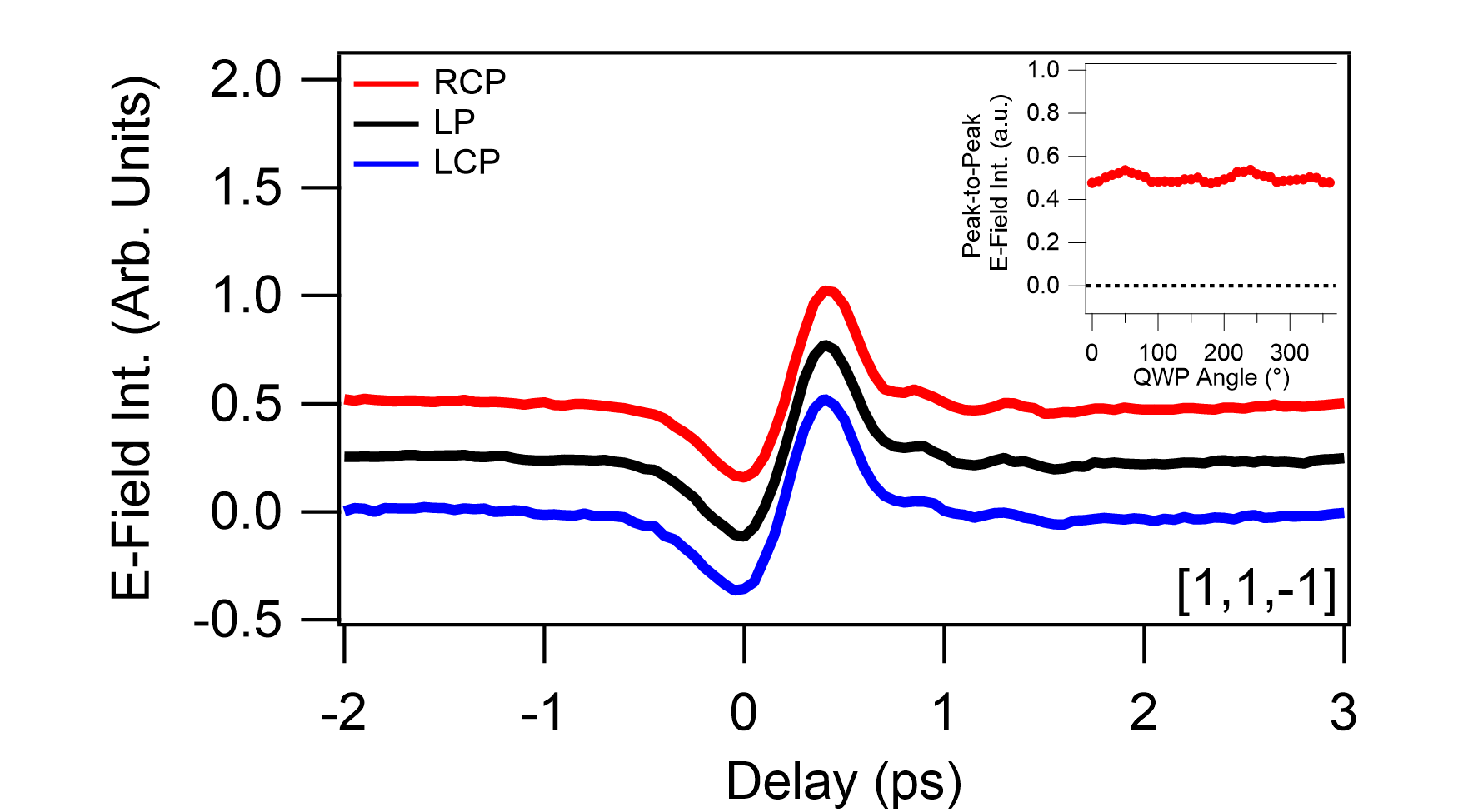}
    \caption{THz emission spectra measured along the [1,1,-1] axis generated by right circular, linear, and left circularly polarized optical pulses (traces are offset for clarity). The inset shows the peak-to-peak E-field amplitude plotted as a function of $\alpha$.}
    \label{fig:cAxis_polDep}
\end{figure}

\begin{figure}[b]
    \centering     
    \includegraphics[width=0.85\columnwidth]{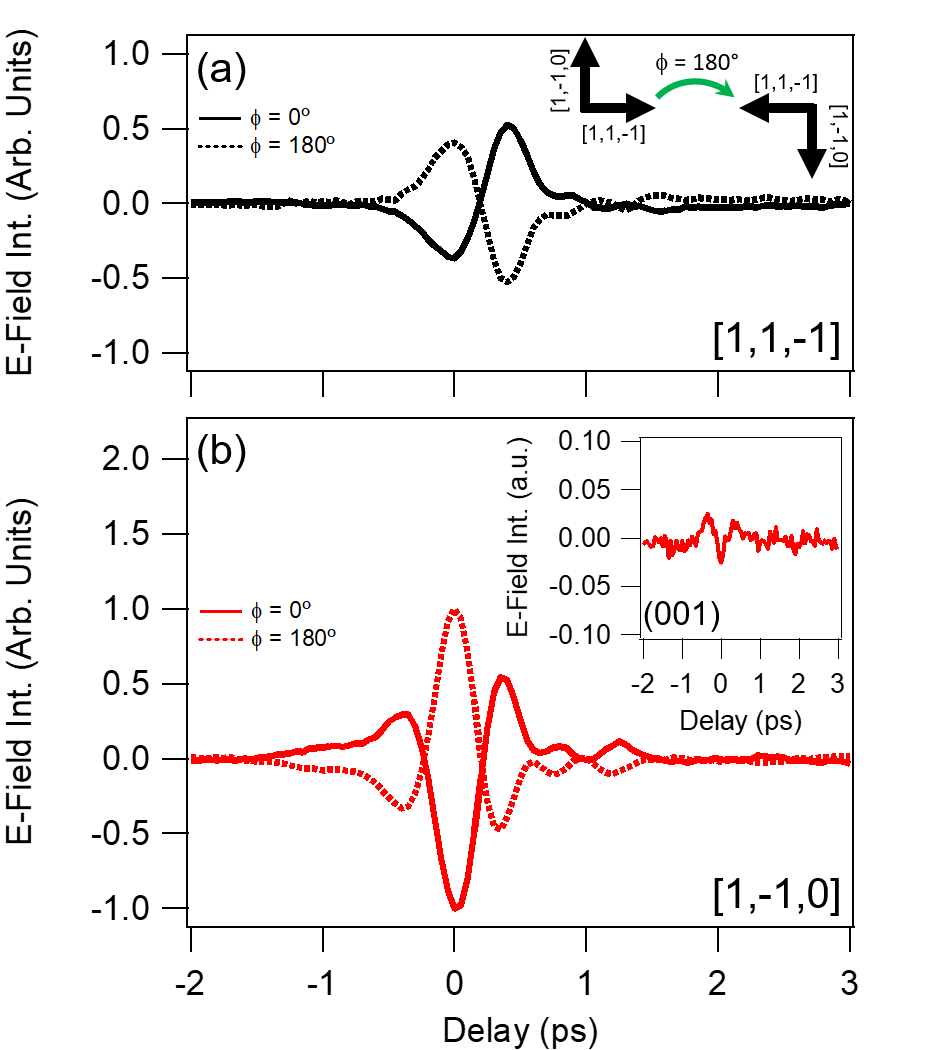}
    \caption{THz emission spectra measured along (a) [1,1,-1] and (b) [1,-1,0], generated from linearly or right circularly polarized light, respectively. Spectra denoted by dashed lines were obtained following a $180^\circ$ azimuthal rotation of the TaAs crystal about the (112) normal. The inset in (b) illustrates the helicity-dependent THz waveform emitted along the [1,-1,0] direction from the (001) surface at normal incidence.}
    \label{fig:Symmetry}
\end{figure}

 In contrast, THz emission polarized along the [1,1,-1] direction, obtained under the same excitation conditions as above, was found to be largely insensitive to the polarization of incident light, and approximately half as intense as that measured along [1,-1,0]. As shown in Fig. \ref{fig:cAxis_polDep}, no variation in the THz waveform and only a small ($<10\%$) variation in the \textbf{E}-field amplitude is observed with rotation of either a \textlambda /4 or a \textlambda /2 waveplate (Supplementary Fig. 4) \cite{Supp}. Fitting the peak-to-peak amplitude of the THz \textbf{E}-field along [1,1,-1] with Eq. (\ref{Equation_1}) shows that the dominant ($\sim90\%$) contribution derives from \textit{D}, as expected by the large offset shown in the inset of Fig. \ref{fig:cAxis_polDep}. Despite this polarization insensitivity, the emitted THz radiation is linearly polarized along the [1,1,-1] axis and exhibits a well-defined directionality. This is illustrated by both an azimuthal dependence that shows the amplitude of the helicity-independent THz waveform to peak along this high symmetry direction (Supplementary Fig. 5)\cite{Supp}, as well as a switching of the emitted THz polarity under a $180^\circ$ rotation of the crystal (Fig. \ref{fig:Symmetry}(a)).
 
 Similarly, helicity-dependent THz radiation polarized along the [1,-1,0] axis exhibits a reversal of polarity under a $180^\circ$ rotation of the sample (Fig. \ref{fig:Symmetry}(b)). This shows that the emitted THz radiation is highly directional; however, unlike Fig. \ref{fig:Symmetry}(a), the directionality of the THz waveform along this axis is determined by the relative orientation that the optical generation pulse makes with the polar $c$-axis. This is most clearly demonstrated by measuring THz emission along the same [1,-1,0] high symmetry direction, but on the (001) face of the crystal, where the $c$-axis lies parallel to the surface normal. Here, the THz pulse emitted at normal incidence is more than 40 times weaker than that measured from the (112) surface under the same conditions (Fig. \ref{fig:Symmetry}(b) inset). However, when repeating the experiment on the (001) face at a $45^\circ$ angle of incidence, the helicity-dependent THz emission is recovered and qualitatively similar to that found from the (112) face (Supplementary Fig. 6) \cite{Supp}.

\begin{figure}[t]
    \centering     
    \includegraphics[width=0.95\columnwidth]{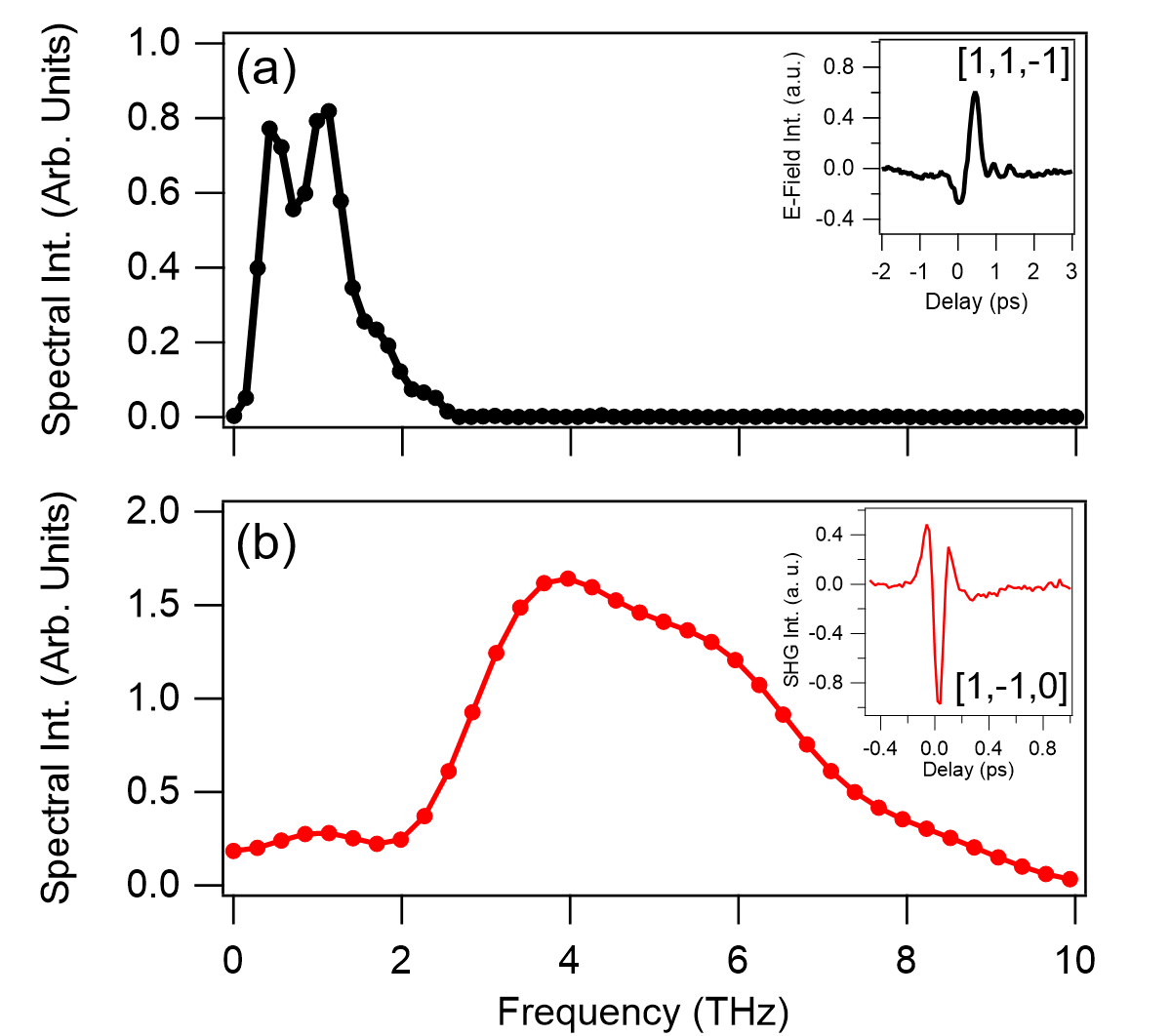}
    \caption{Intensity power spectra obtained by Fourier transforming (a) time-domain THz waveforms measured along [1,1,-1] using free space EOS (inset) and (b) helicity-and-time-dependent THz pulses measured with TFISH (inset) following appropriate subtraction of the time-resolved SHG dynamics along the [1,-1,0] axis.}
    \label{fig:EOS_TFISH_BW}
\end{figure}

 THz pulses emitted along both high symmetry axes of the (112) surface are found to scale linearly with laser fluence and exhibit no change in waveform or frequency content as higher excitation fluences are used (Supplementary Fig. 7) \cite{Supp}. By Fourier transforming the THz time-domain traces shown above, one finds the spectral weight of the THz intensity power spectrum along [1,1,-1] to be shifted towards lower frequencies ($\sim1.0$ THz) (Fig. \ref{fig:EOS_TFISH_BW} (a)), and thus longer timescales, as compared to that of the helicity-dependent THz radiation emitted along [1,-1,0] (Supplementary Fig. 7)\cite{Supp}. As it turns out, spectra measured along [1,-1,0] by free space EOS are limited by the detection bandwidth of the $<110>$ ZnTe crystal (Supplementary Fig. 8) \cite{PrasankumarBook,Supp}. This remains true even when thinner ZnTe crystals are used, making it difficult to accurately estimate the emitted THz bandwidth using this technique.  
 
 To provide a better estimate for the bandwidth of the helicity-dependent THz pulse polarized along [1,-1,0], we used TFISH \cite{Nahata1998}. Here, the sensitivity of optical second harmonic generation (SHG) to broken inversion symmetry enables us to detect the electric field of the transient THz pulse without the bandwidth limitations imposed by the strong vibrational resonances in electro-optic crystals. More specifically, the THz field emitted after exciting the sample with a circularly polarized 800 nm pump pulse (as in the experiments described above) induces a change in the SHG signal polarized along the in-plane [1,-1,0] direction. This can be measured with a separate probe beam through a \textchi\textsuperscript{(3)} process that acts in addition to the usual \textchi\textsuperscript{(2)} component. In this way, TFISH can be described as a four-wave mixing process in which light of frequency $2\omega$ is generated from mixing light with frequencies $\omega$, $\omega$, and $\omega_{THz}$, and represented by the second order electric polarization,  
 \begin{equation}
     P_{i}(2\omega) = (\chi_{ijk}^{(2)}+\chi_{ijkl}^{(3)}E_{l}(\omega_{THz}))E_{j}(\omega)E_{k}(\omega),
     \label{Equation_2}
 \end{equation}
 where $\chi^{(3)}$ has the same symmetry constraints as $\chi^{(2)}$, leaving the symmetry of the SHG pattern unchanged.
 
 Time-domain waveforms obtained from our TFISH measurements are shown in the inset of Fig. \ref{fig:EOS_TFISH_BW} (b), where the emitted THz pulse is isolated after subtracting the longer time dynamics associated with the pump-induced change in the SHG signal (Supplementary Fig. 9) \cite{Supp}. As compared to the THz waveform measured by EOS, the temporal duration of the emitted THz pulse detected by TFISH is significantly shorter, with an intensity power spectrum that yields bandwidth out to 10 THz (Fig. \ref{fig:EOS_TFISH_BW} (b)). While substantially broader than that obtained by free space EOS, even this is limited by the temporal resolution of the time-resolved SHG experiment, meaning that an upper limit of $\sim100$ fs can be placed on the underlying dynamics responsible for the helicity-dependent THz radiation emitted from the TMMP family of WSM.
 
 From the data presented above, we can conclude that the helicity-dependent THz emission shown in Fig. \ref{fig:CPGE} derives from an ultrafast photocurrent flowing along the [1,-1,0] high symmetry direction. This finding is consistent with the previously reported CPGE in these materials \cite{Ma2017,Kai2017}, and is further supported by symmetry considerations, outlined in supplementary section X \cite{Supp}. In particular, for circularly polarized light, denoted by the complex E-field $E$ and $E^{*}=E(-k,-\omega)$, normally incident on the (112) face, symmetry constraints placed on the CPGE response tensor, $\gamma_{ls}$, by the $C_{4v}$ point group of the crystal allow for a helicity-dependent, transverse photocurrent ($J$) to flow along the [1,-1,0] axis, while symmetry forbids a helicity-dependent photocurrent along [1,1,-1]:
 \begin{equation}
    \begin{split}
     & J_{[1,-1,0]}^{CPGE} = i\frac{\gamma_{xy}}{\sqrt{3}}(\vec{E}\times\vec{E}^{*})_{[1,1,2]}\\
     & J_{[1,1,-1]}^{CPGE} = 0.
     \label{Eqn:CPGE_112_MT}
     \end{split}
 \end{equation}
 Furthermore, as expected from Fig \ref{fig:Symmetry}(b) the in-plane photocurrent, $J_{[1,-1,0]}^{CPGE}$, will necessarily switch sign following a $180^\circ$ rotation of the crystal, while photocurrent generation from light normally incident on the (001) face is found to be symmetry forbidden (Supplementary Fig. 6) \cite{Supp}. Hence, our experimental findings are in complete agreement with what is expected by symmetry for the CPGE. However, before assigning the underlying mechanism of the helicity-dependent photocurrent to this effect, it is important to note that such photocurrents can also arise from alternate mechanisms, including the circular photon drag (CPDE) and spin-galvanic effects (SGE)  \cite{Ganichev2003,Ganichev2002}. Therefore, to address the role played by these additional effects, we consider further the symmetry constraints imposed on the CPDE tensor as well as the dynamical insight into the ultrafast photocurrent gained through our THz emission measurements. 
 
 While typically responsible for longitudinal photocurrents contained within the scattering plane, transverse helicity-dependent photocurrents, such as those observed here, can result from the CPDE \cite{Shalygin2016} and are allowed under the $C_{4v}$ symmetry of TaAs, as indicated by the independent tensor elements $T_{xyxy}$ and $T_{xzxz}$ of the CPDE response tensor \cite{Supp}. However, in contrast to the CPGE, circularly polarized light normally incident to the (112) face will impart a momentum $q$ along the [1,1,2] normal, generating helicity-dependent photocurrents along the [1,1,-1] axis as opposed to the [1,-1,0] direction:
\begin{equation}
    \begin{split}
     J_{[1,-1,0]}^{CPDE} & = 0\\
     J_{[1,1,-1]}^{CPDE} & = i \frac{2}{3\sqrt{3}}(T_{xyxy}+T_{xzxz})q_{[1,1,2]}(\vec{E}\times\vec{E}^{*})_{[1,1,2]}.
     \label{Eqn:CPDE_112_MT}
     \end{split}
\end{equation}
 In other words, helicity-dependent photocurrents generated by either the CPGE or CPDE in TaAs can be distinguished from one another based on the direction that current flows relative to the orthogonal high symmetry axes of the (112) surface. Experimentally, Fig. \ref{fig:cAxis_polDep} shows that THz emission along [1,1,-1] is largely polarization independent, with only a small helicity-dependent contribution, as found from a fit of the peak-to-peak THz amplitude as a function of QWP angle (Supplementary Fig. 4(a))\cite{Supp}. While this small helicity-dependence to the [1,1,-1] photocurrent most likely originates from the CPDE, symmetry forbids such a mechanism from generating the dominant ultrafast helicity-dependent photocurrent along [1,-1,0]. For this reason, we exclude the CPDE as an origin for the helicity-dependent photocurrent seen in Fig \ref{fig:CPGE}.

 In contrast, distinguishing between a helicity-dependent photocurrent arising from the CPGE versus the SGE requires dynamical insights that can be gained by analysis of the THz waveform. Unlike a non-resonant second order process, the helicity-dependent THz radiation emitted here corresponds to a real, transient current. Consequently, the spectral bandwidth and waveform of the emitted THz pulse are not dependent on that of the excitation pulse, but are intrinsic features of the ultrafast current generated in these materials \cite{THzSpecBook}. For a pulsed excitation, the decay of the helicity-dependent photocurrent will be determined by either the momentum or spin relaxation time, depending upon whether it originates from the CPGE or the SGE \cite{Ganichev2003}. The broad emission bandwidth observed along the [1,-1,0] axis (Fig. \ref{fig:EOS_TFISH_BW}(b)) implies a lifetime of $< 100$ fs for the excited photocurrent. This is more consistent with a current decay following the momentum relaxation time of a free carrier than a slower spin relaxation due to asymmetric spin-flip scattering of photoexcited carriers \cite{Ganichev2003,Ganichev2002}. When coupled with the above symmetry analysis, this leaves the most likely origin of the helicity-dependent photocurrent to be injection photocurrents that give rise to the CPGE. 
 
 As compared to the helicity-dependent THz emission observed along the [1,-1,0] axis, the fundamental mechanism underlying THz emission polarized along [1,1,-1] is distinct. Since this axis contains a projection of the inversion symmetry-broken $c$-axis, both the polarization independence and the well-defined directionality of the photocurrent suggest an underlying mechanism rooted in broken inversion symmetry. As a result, THz emission measured along the [1,1,-1] axis of the (112) surface is intrinsic to the non-centrosymmetric crystal structure of TaAs and can likely be understood as an optical excitation producing electron-hole pairs, regardless of polarization, which are then separated by the dipole-like field of the polar Ta-As bond lying along the $c$-axis. Such a microscopic picture is consistent with that of a shift current, believed to be the origin of both the giant anisotropic second harmonic signal\cite{Wu2016,Patankar2018} and the colossal bulk photovoltaic effect\cite{Osterhoudt2017} seen in these materials.
 
 In this regard, despite our use of femtosecond optical pulses whose energy is well above the energy scale associated with the Weyl cone, the THz emission spectra shown here exhibit the same fundamental behavior as observed in static photocurrent experiments. Despite this similarity, assigning a microscopic mechanism to the ultrafast photocurrents observed in TaAs becomes challenging, as arguments rooted in Weyl physics hold for mid-IR excitation\cite{Ma2017,Osterhoudt2017} but not for optical excitation, where details of the trivial band structure are expected to become more relevant \cite{Lee2015}. Rather, our findings suggest that under optical excitation these transient photocurrents are intrinsic to the underlying crystal symmetry of TaAs, whose $C_{4v}$ symmetry belongs to the gyrotropic crystal class, and may not have an explicit link to Weyl physics beyond the fact that such a symmetry supports the existence of Weyl nodes in the electronic structure.
 
\textit{Conclusion:} 
 In closing, we performed THz emission spectroscopy on the (112) and (001) surfaces of the TMMP WSM TaAs. Our data enables us to clearly distinguish between helicity-dependent photocurrents generated within the $ab$-plane and polarization-independent photocurrents flowing along the non-centrosymmetric $c$-axis. Such findings are in excellent agreement with previous static photocurrent measurements. However, by considering both the physical constraints imposed by symmetry and the temporal dynamics intrinsic to current generation and decay, we can attribute these transient photocurrents to the underlying crystal symmetry of the TMMP family of WSM.
 
\textit{Acknowledgements:} 
 This work was performed at the Center for Integrated Nanotechnologies at Los Alamos National Laboratory (LANL), a U.S. Department of Energy, Office of Basic Energy Sciences user facility.  We gratefully acknowledge the support of the U.S. Department of Energy through the LANL LDRD Program, the G. T. Seaborg Institute, and the Center for Advancement of Topological Semimetals, an Energy Frontier Research Center funded by the U.S. Department of Energy Office of Science, Office of Basic Energy Sciences, through the Ames Laboratory under its Contract No. DE-AC02-07CH11358. RIT was the 2017-2018 Los Alamos National Laboratory Rosen Scholar, supported by LDRD $\#$20180661ER.

\bibliography{Sirica}

\end{document}


\title{Supplementary Information: Tracking ultrafast photocurrents in the Weyl semimetal TaAs using THz emission spectroscopy}

\author{N. Sirica}
\email{nsirica@lanl.gov}
\affiliation{Center for Integrated Nanotechnologies, Los Alamos National Laboratory, Los Alamos, NM 87545, USA}
\author{R.I. Tobey}
\affiliation{Center for Integrated Nanotechnologies, Los Alamos National Laboratory, Los Alamos, NM 87545, USA}
\affiliation{Zernike Institute for Advanced Materials, University of Groningen, Groningen, The Netherlands}
\author{L.X. Zhao}
\affiliation{Institute of Physics, Chinese Academy of Sciences, Beijing 100190, China}
\author{G.F. Chen}
\affiliation{Institute of Physics, Chinese Academy of Sciences, Beijing 100190, China}
\author{B. Xu}
\affiliation{Institute of Physics, Chinese Academy of Sciences, Beijing 100190, China}
\author{R. Yang}
\affiliation{Institute of Physics, Chinese Academy of Sciences, Beijing 100190, China}
\author{B. Shen}
\affiliation{Department of Physics and Astronomy, University of California, Los Angeles, CA 90095, USA}
\author{D.A. Yarotski}
\affiliation{Center for Integrated Nanotechnologies, Los Alamos National Laboratory, Los Alamos, NM 87545, USA}
\author{P. Bowlan}
\affiliation{Center for Integrated Nanotechnologies, Los Alamos National Laboratory, Los Alamos, NM 87545, USA}
\author{S.A. Trugman}
\affiliation{Center for Integrated Nanotechnologies, Los Alamos National Laboratory, Los Alamos, NM 87545, USA}
\author{J.X. Zhu}
\affiliation{Center for Integrated Nanotechnologies, Los Alamos National Laboratory, Los Alamos, NM 87545, USA}
\author{Y.M. Dai}
\affiliation{Center for Integrated Nanotechnologies, Los Alamos National Laboratory, Los Alamos, NM 87545, USA}
\affiliation{School of Physics, Nanjing University, Nanjing 210093, China}
\author{A.K. Azad}
\affiliation{Center for Integrated Nanotechnologies, Los Alamos National Laboratory, Los Alamos, NM 87545, USA}
\author{N. Ni}
\affiliation{Department of Physics and Astronomy, University of California, Los Angeles, CA 90095, USA}
\author{X.G. Qiu}
\affiliation{Institute of Physics, Chinese Academy of Sciences, Beijing 100190, China}
\author{A.J. Taylor}
\affiliation{Center for Integrated Nanotechnologies, Los Alamos National Laboratory, Los Alamos, NM 87545, USA}
\author{R.P. Prasankumar}
\email{rpprasan@lanl.gov}
\affiliation{Center for Integrated Nanotechnologies, Los Alamos National Laboratory, Los Alamos, NM 87545, USA}


\maketitle
\section{Comparison of terahertz emission spectra from tantalum arsenide and niobium arsenide}

\begin{figure}[h]
    \centering     
    \includegraphics[width=\columnwidth]{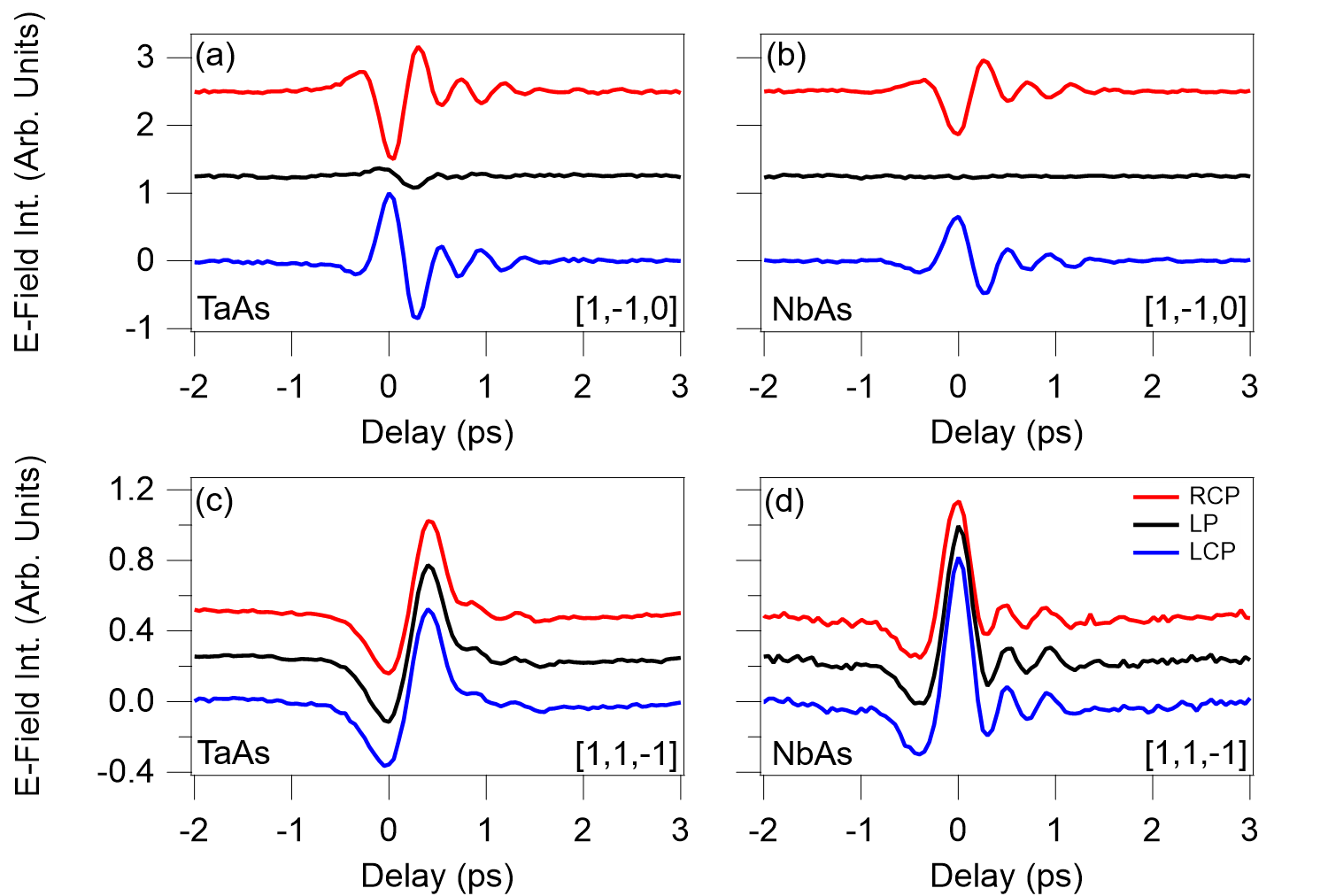}
    \caption{Comparison of THz emission spectra measured along the [1,-1,0] and [1,1,-1] axes generated by right circular, linear, and left circularly polarized femtosecond optical pulses incident on the (112) face of TaAs ((a) and (c)) and NbAs ((b) and (d).}
    \label{fig:TaAsvsNbAs}
\end{figure}

\newpage
\section{Experimental Geometry and Crystal Structure}

\begin{figure}[h]
    \centering     
    \includegraphics[width=\columnwidth]{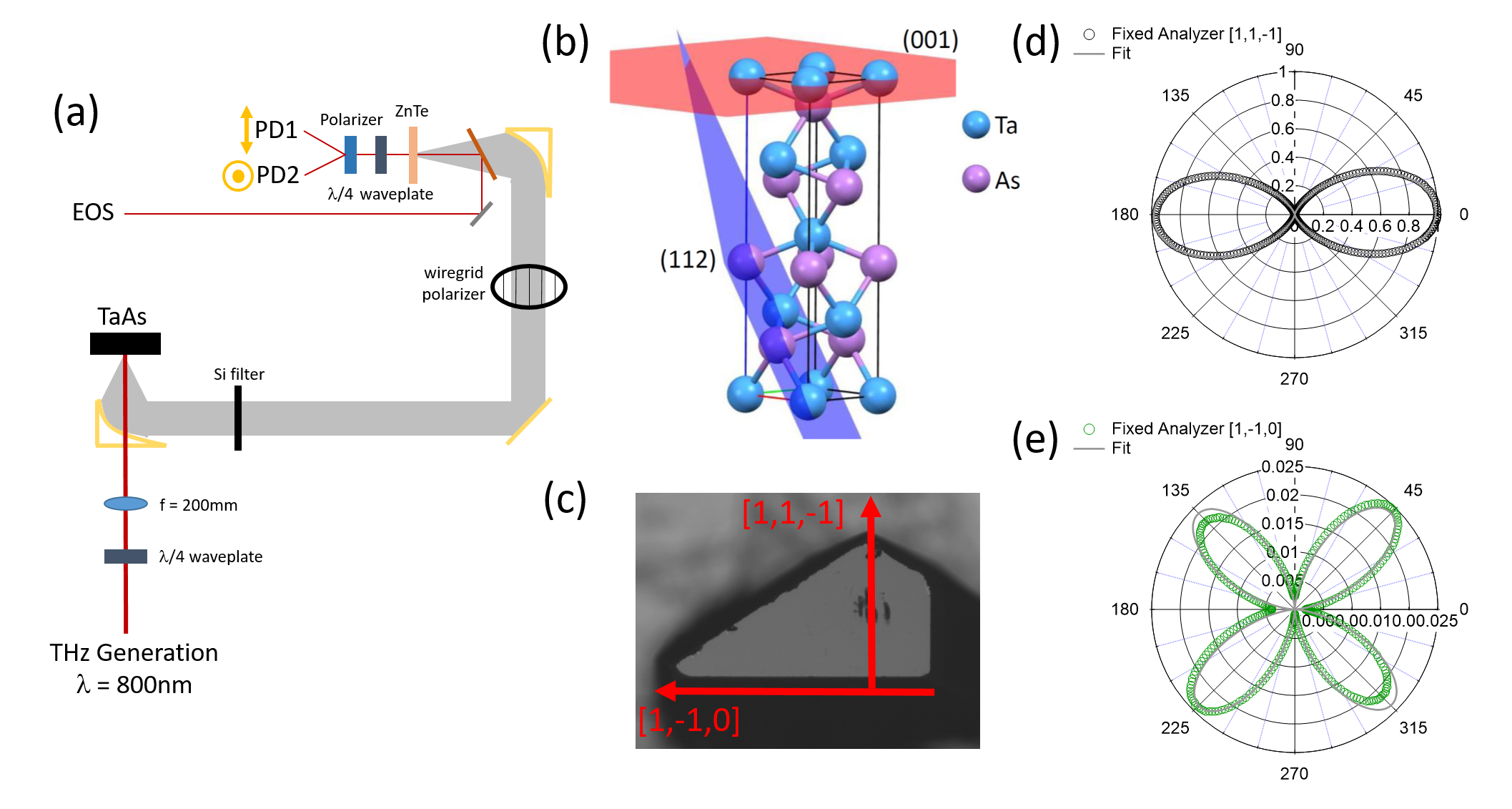}
    \caption{(a) Sketch of the near normal incidence geometry described in the \textit{Experimental} section of the main text. Here, specularly emitted THz pulses polarized parallel to the scattering plane were selected by a wire grid polarizer and detected through free space electro-optic sampling in a ZnTe crystal. The THz polarization was measured relative to the high symmetry axes of the (001) and (112) planes of the tetragonal unit cell shown in (b). On the (112) face, the two in-plane high symmetry axes labeled in (c) were determined by Laue diffraction as well as from static second harmonic generation (SHG) measurements made along the (d) [1,1,-1] and (e) [1,-1,0] directions. A fit of the SHG pattern shows the crystallographic point group to be well described by a $C_{4v}$ symmetry, in agreement with previous measurements \cite{Wu2016}.}
    \label{fig:HWPDep_abplane}
\end{figure}

\newpage
\section{Half waveplate dependence of the terahertz emission spectra measured along [1,-1,0]}

\begin{figure}[h]
    \centering     
    \includegraphics[width=\columnwidth]{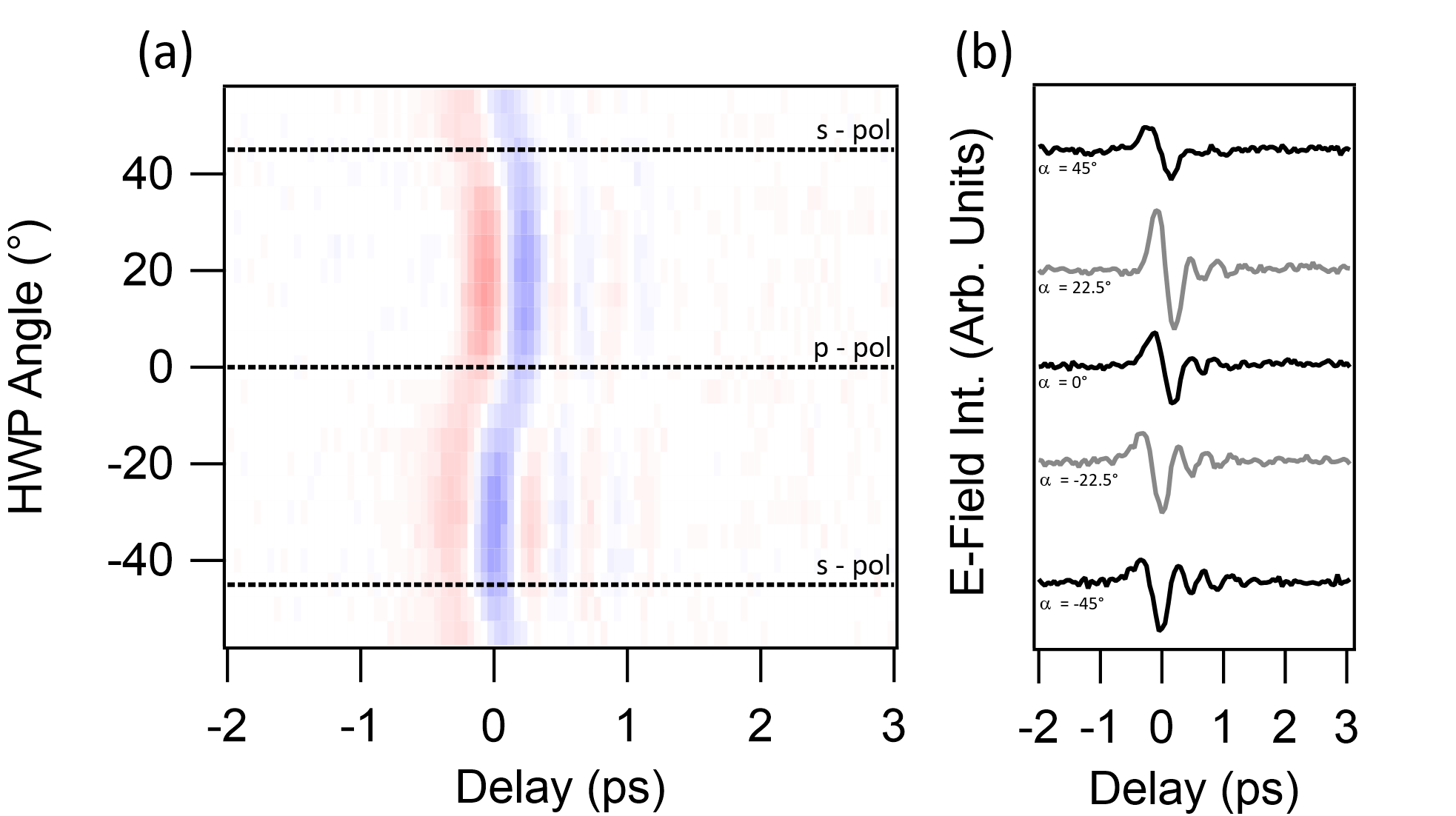}
    \caption{(a) False color plot and (b) select time-dependent THz traces, illustrating changes in the emitted THz waveform upon changing the polarization of the optical generation pulse from horizontal ($p$) to vertical ($s$). Traces shown in (b) were obtained using half waveplate angles of $\pm45^\circ$, $\pm22.5^\circ$, and $0^\circ$. While a change in amplitude and phase of the emitted THz pulse is evident, normalizing the color scale in (a) with respect to Fig. 1 (a) of the main text shows that emission along [1,-1,0] is considerably weaker for linear versus circularly polarized optical generation pulses.}
    \label{fig:HWPDep_abplane}
\end{figure}

\newpage
\section{Polarization dependence of the emitted terahertz amplitude along [1,1,-1]}

\begin{figure}[h]
    \centering     
    \includegraphics[width=\columnwidth]{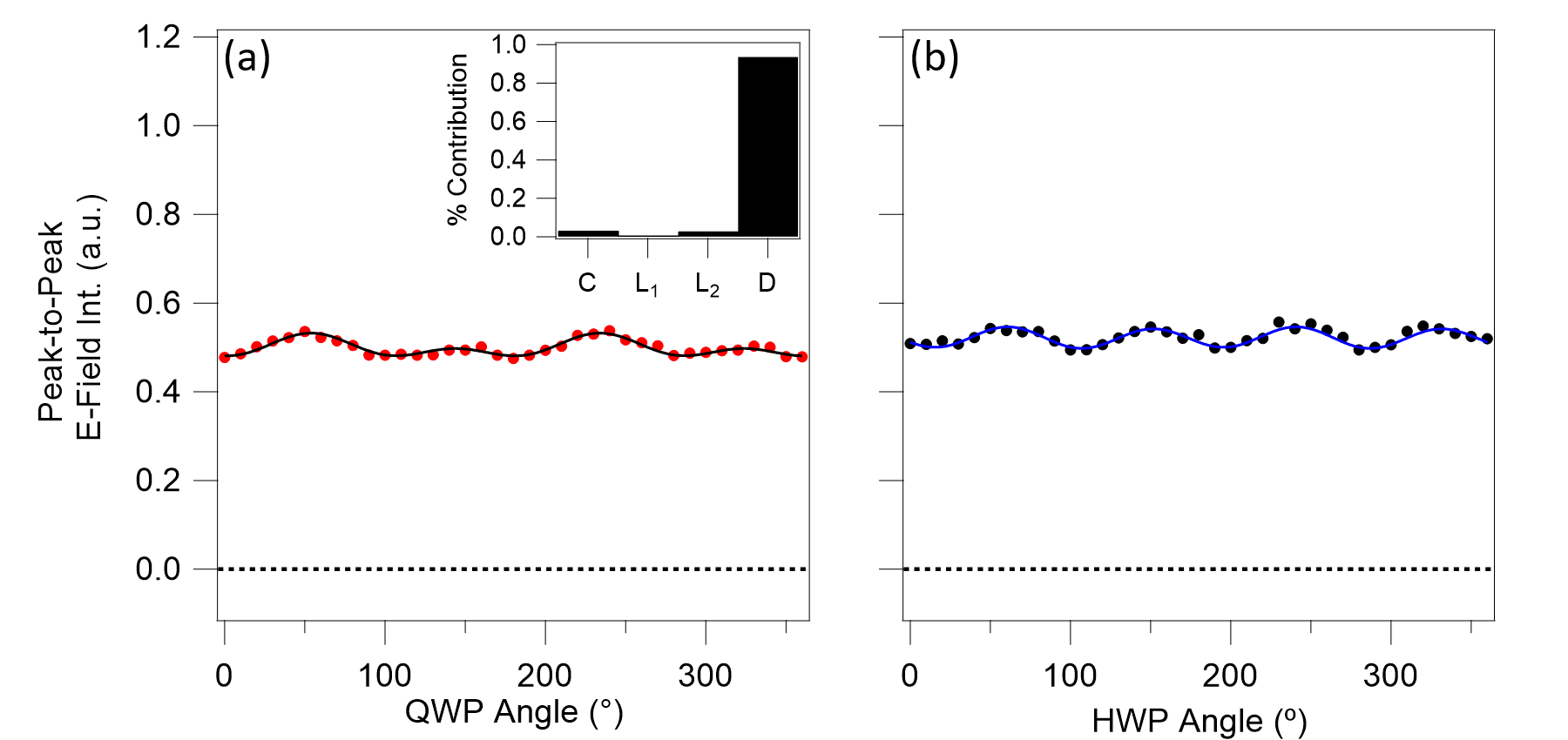}
    \caption{Peak-to-peak E-field intensity of the emitted THz waveform polarized along [1,1,-1] and plotted as a function of (a) quarter waveplate and (b) half waveplate angles. Fitting our data with the expression in Eq. (1) of the manuscript (inset of (a)) reveals the dominant contribution to be from the polarization-independent offset, \textit{D}.}
    \label{fig:Pk2Pk_cAxis}
\end{figure}

\newpage
\section{Azimuthal dependence}

\begin{figure}[h]
    \centering     
    \includegraphics[width=0.6\columnwidth]{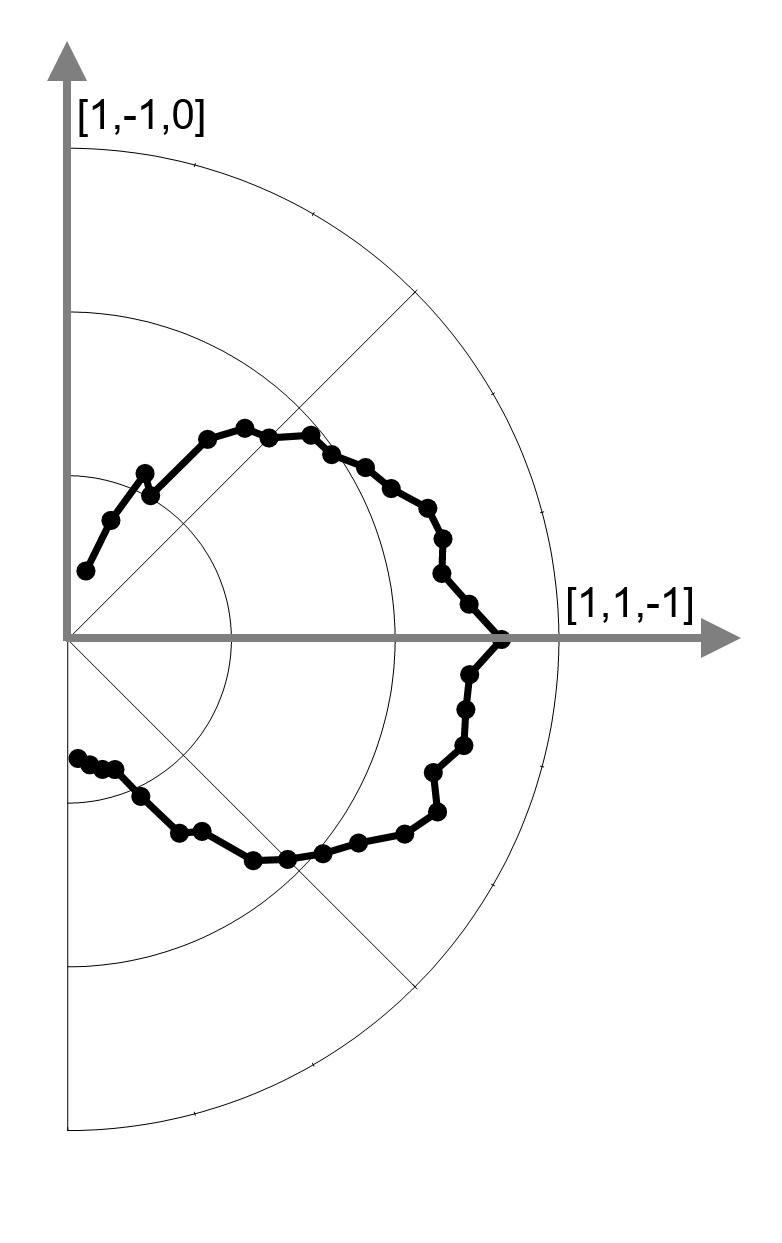}
    \caption{Azimuthal dependence of the peak-to-peak E-field intensity for helicity-independent THz pulses emitted from the (112) face of NbAs. Here, the crystal is rotated about the (112) normal while keeping the linear polarizations of the generating optical pulse and detected THz pulse constant. The emitted THz amplitude peaks when the polarization of the optical generation pulse is parallel to the [1,1,-1] axis of the crystal.}
    \label{fig:Azimuth}
\end{figure}

\newpage
\section{Helicity-dependent terahertz emission from the (001) face}

\begin{figure}[h]
    \centering     
    \includegraphics[width=\columnwidth]{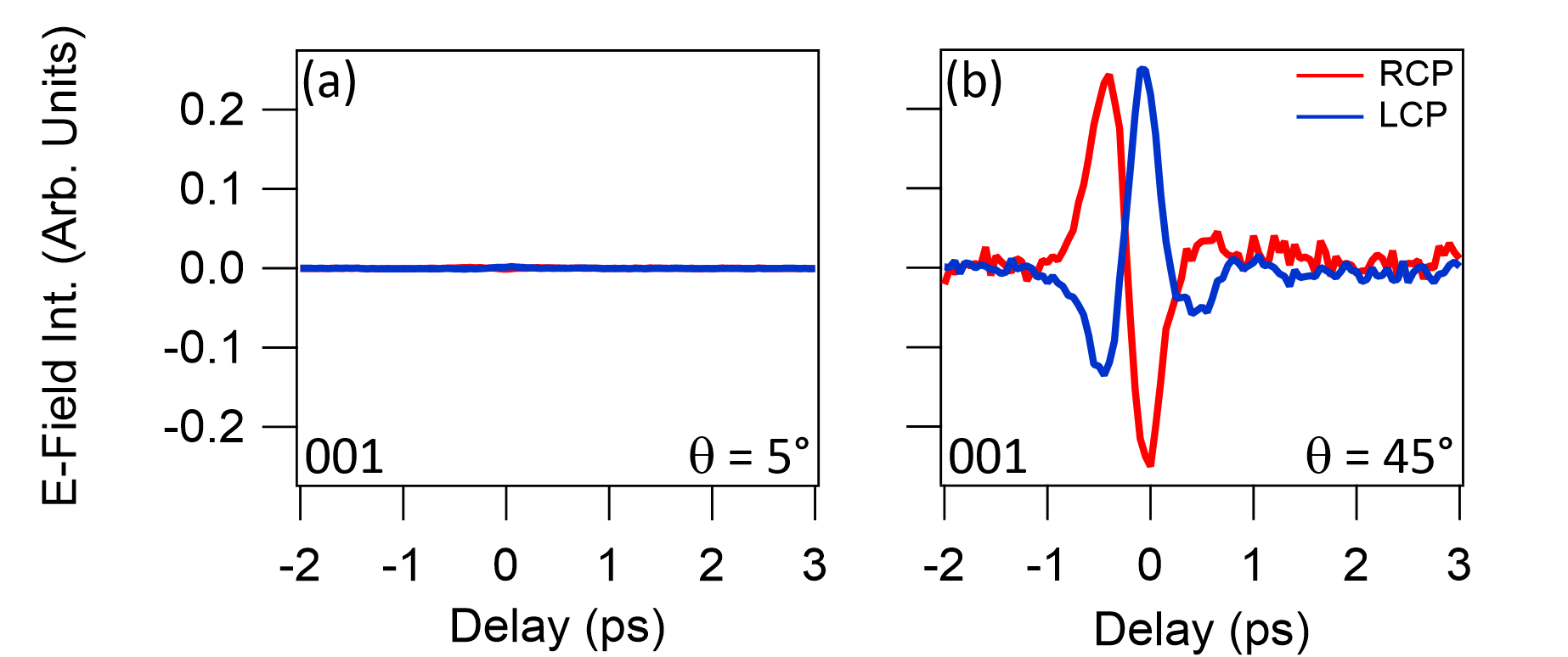}
    \caption{Helicity-dependent THz emission taken at (a) $5^\circ$ and at (b) $45^\circ$ angles of incidence along the [1,-1,0] axis of the (001) face. The emitted THz pulses are polarized perpendicular to the scattering plane and waveform amplitudes are normalized with respect to that emitted along the same axis of the (112) surface.}
    \label{fig:CPGE_001Face}
\end{figure}

\newpage
\section{Fluence dependence and intensity power spectrum}

\begin{figure}[h]
    \centering     
    \includegraphics[width=0.8\columnwidth]{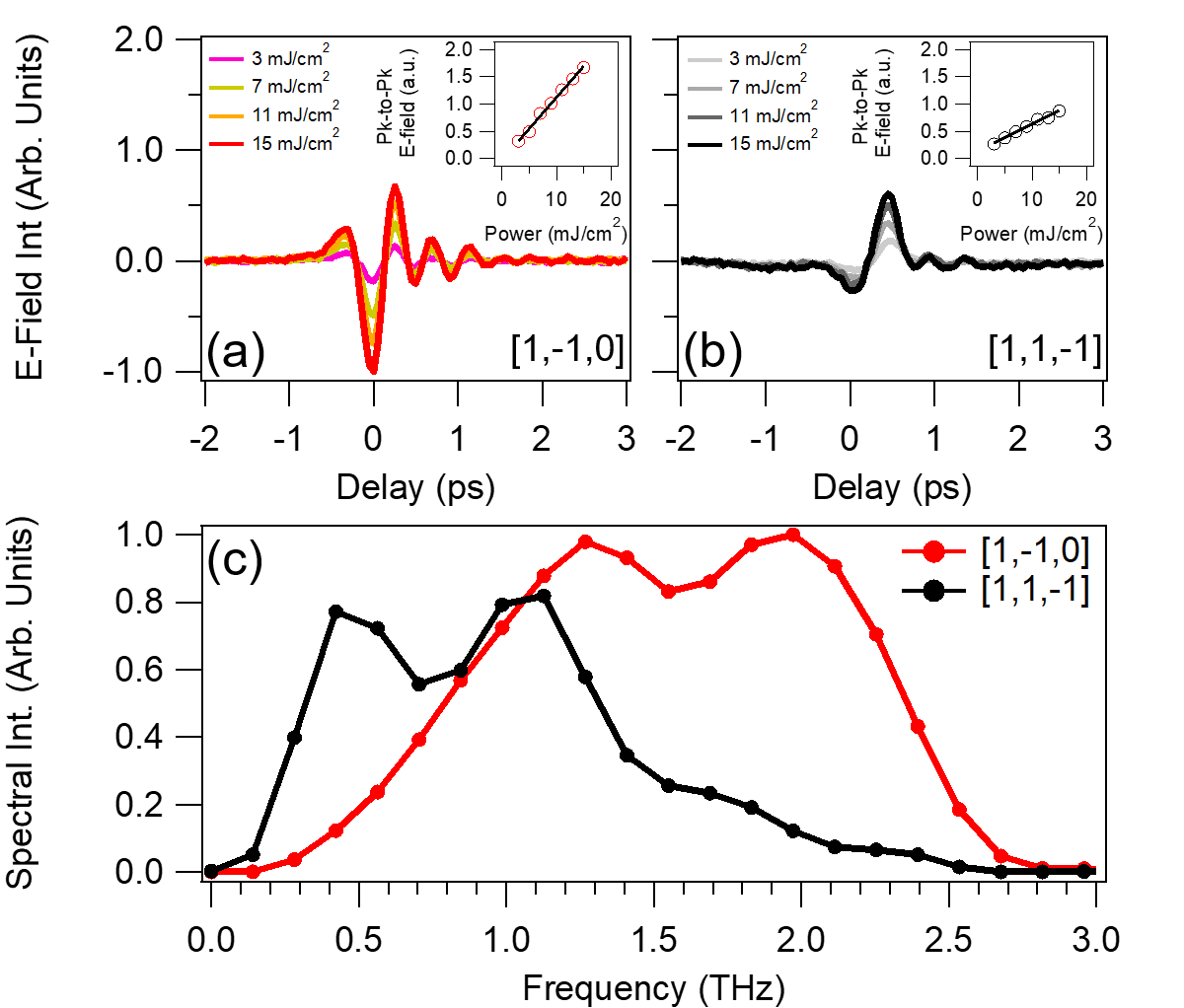}
    \caption{Fluence dependence of (a) THz pulses polarized along [1,-1,0], generated from right circularly polarized light and (b) THz pulses polarized along [1,1,-1], generated from linearly polarized light. Insets denote the linear scaling of the peak-to-peak THz E-field as a function of excitation fluence. Note the measured THz \textbf{E}-field increases more gradually with fluence for the polarization-independent THz pulse emitted along [1,1,-1] as compared to that emitted along [1,-1,0]. (c) Intensity power spectra obtained by Fourier transforming the time-domain THz spectra measured along [1,-1,0] and [1,1,-1] using free space EOS. The dip in the power spectra at $\sim1.7$ THz is common to both TaAs and ZnTe (Supplementary Fig. 8)}
    \label{fig:Fluence}
\end{figure}

\newpage
\section{Intensity power spectrum generated from zinc telluride}

\begin{figure}[h]
    \centering     
    \includegraphics[width=\columnwidth]{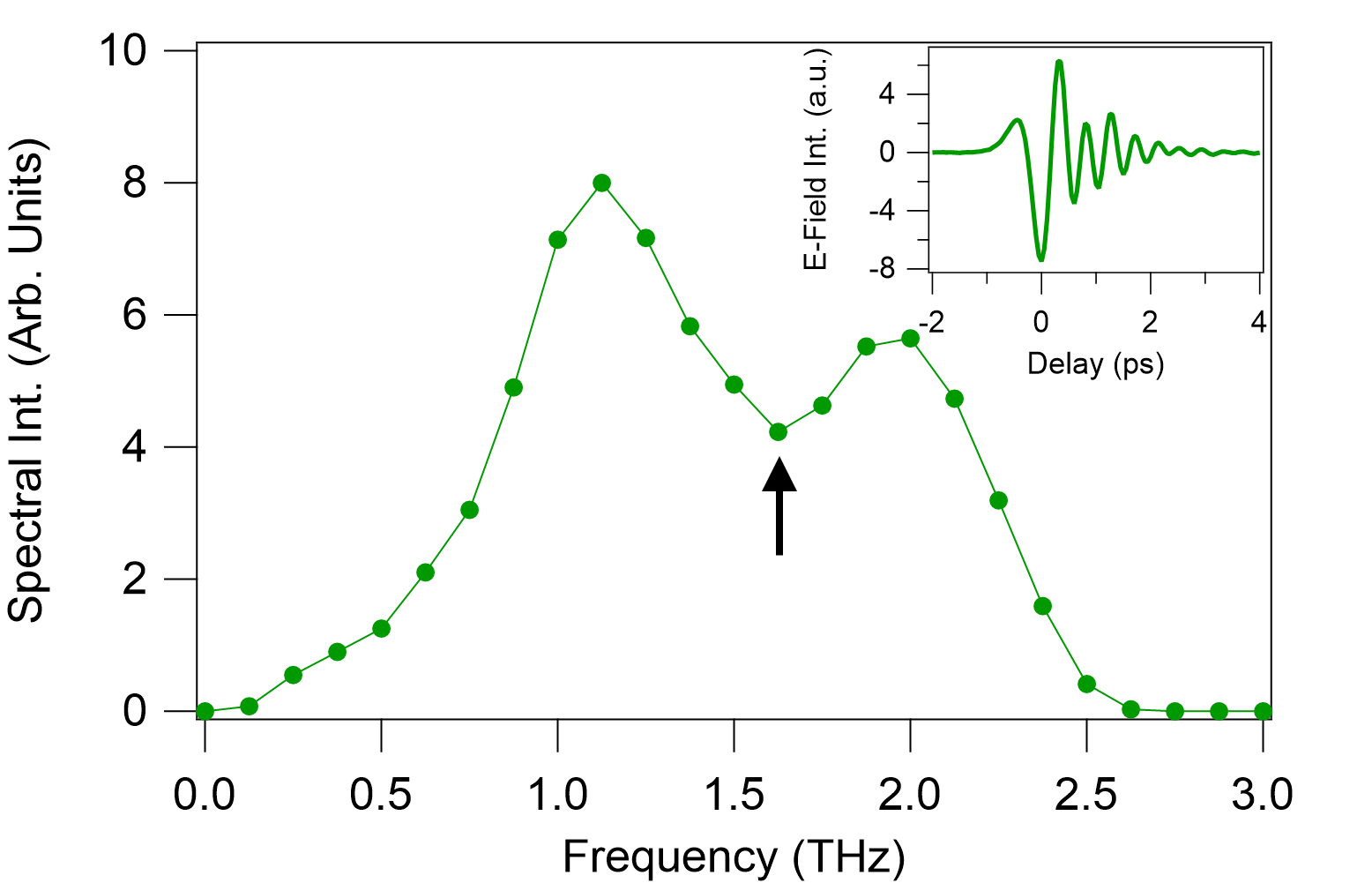}
    \caption{Intensity power spectrum obtained by Fourier transforming the time domain THz spectra emitted from a 0.5 mm thick $<110>$ ZnTe crystal (inset). THz pulses were generated in a transmission geometry under similar excitation conditions to that used for TaAs. Note that the prominent dip at $\sim1.7$ THz in the power spectrum is common to both ZnTe and TaAs.}
    \label{fig:ZnTe}
\end{figure}

\newpage
\section{TFISH background subtraction}

\begin{figure}[h]
    \centering     
    \includegraphics[width=\columnwidth]{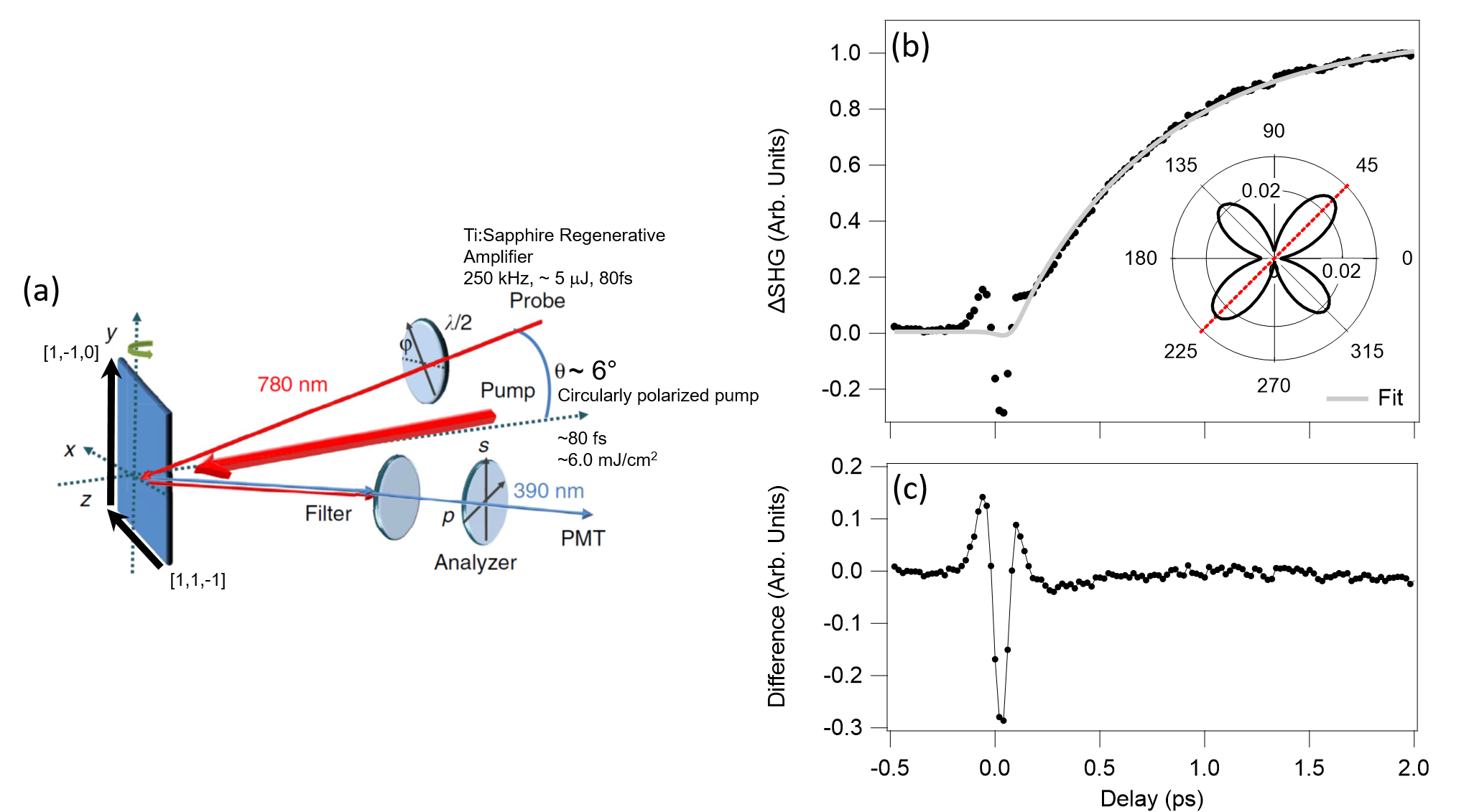}
    \caption{(a) Sketch of the experimental setup used for measuring a TFISH signal \cite{Sheu2014}. Here, circularly polarized light normally incident on the (112) face of TaAs generates a transient photocurrent along the [1,-1,0] axis. By probing the SHG signal polarized along this direction, time-dependent changes in the static SHG signal due to both optical excitation and the E-field of the emitted THz pulse can be detected, in accordance with equation (2) of the main text. From the time-resolved SHG trace shown in (b), a single cycle waveform of the emitted THz pulse can be obtained in (c), following appropriate subtraction of the exponential rise associated with the TR-SHG dynamics. The inset of (b) shows a polar plot of the SHG signal vs. incident polarization, and the red line denotes the polarization for which the time-resolved SHG trace was measured.}
    \label{fig:TFISH_BkgdSub}
\end{figure}



\newpage
\section{Symmetry Analysis}
The linear scaling of the peak-to-peak THz E-field as a function of excitation fluence shown in Fig. \ref{fig:Fluence} of the supplementary information indicates that the ultrafast photocurrent originates from a second order process and therefore necessitates that inversion symmetry be broken. In such a case, the creation of a DC current density ($J_{l}^{DC}$) in response to an oscillating optical field $E$ is given by \cite{Ma2017_Supp,Quereda2018,Ganichev2003}
\begin{equation}
     J_{l}^{DC} = \sigma_{ljk}E_{j}E_{k}^{*}.
     \label{Photocurrent}
\end{equation}
This expression remains valid even for the generation of ultrafast photocurrents, as the duration of the emitted THz pulse will be considerably longer than that of the oscillating optical field. By expanding the photocurrent response tensor, $\sigma_{ljk}$, with respect to the wavevector, $q_{n}$, 
\begin{equation}
     J_{l}^{DC} = \sigma_{ljk}(q,\omega)E_{j}E_{k}^{*}=\eta_{ljk}(\omega)E_{j}E_{k}^{*}+T_{lnjk}(\omega)q_{n}E_{j}E_{k}^{*}+...,
     \label{Photocurrent_Expansion}
\end{equation}
the photocurrent can be separated on the basis of a finite momentum transfer imparted by the incident photon. To first order, this includes $\sigma_{ljk}(0,\omega)=\eta_{ljk}(\omega)$, which retains a frequency ($\omega$) dependence but has no dependence on the radiation wavevector, and $T_{lnjk}(\omega)$, which is linear in $q_{n}$. Such terms, characterized by the response tensors $\eta_{ljk}$ and $T_{lnjk}$, are responsible for the photogalvanic and photon drag effects, respectively. 

Considering that the photocurrent density is a real quantity and must therefore remain unchanged under complex conjugation, the real part of $\eta_{ljk}$ and $T_{lnjk}$ must be symmetric under coordinate exchange, while the imaginary part will be antisymmetric. With regards to helicity-dependent photocurrents, it is the imaginary part of the photocurrent response tensor that gives rise to either the circular photogalvanic (CPGE) or the circular photon drag (CPDE) effects \cite{Ganichev2003,Quereda2018}. Hence, under coordinate permutation, the CPGE can be expressed as
\begin{equation}
     \begin{split}
        J_{l}^{CPGE} &= i\eta_{ljk}^{antisym}(E_{j}E_{k}^{*}-E_{k}E_{j}^{*})=i\gamma_{ls}\epsilon_{sjk}E_{j}E_{k}^{*}\\&=i\gamma_{ls}(\vec{E}\times \vec{E^{*}})_{s},
        \label{CPGE_Equation}
     \end{split}
\end{equation}
while the CPDE is given by,
\begin{equation}
    \begin{split}
        J_{l}^{CPDE} & = iT_{lnjk}^{antisym}q_{n}(E_{j}E_{k}^{*}-E_{k}E_{j}^{*})=iT_{lns}q_{n}\epsilon_{sjk}E_{j}E_{k}^{*}\\ &=iT_{lns}q_{n}(\vec{E}\times \vec{E^{*}})_{s}.
     \label{CPDE_Equation}
     \end{split}
\end{equation}
In both cases, the Levi-Civita tensor,$\epsilon_{sjk}$, is used to contract $\eta_{ljk}^{antisym}$ and $T_{lnjk}^{antisym}$ to the $2^{nd}$ and $3^{rd}$ rank tensors $\gamma_{ls}$ and $T_{lns}$, which define the CPGE and CPDE response tensors, respectively. 

\begin{figure}[t]
    \centering     
    \includegraphics[width=0.55\columnwidth]{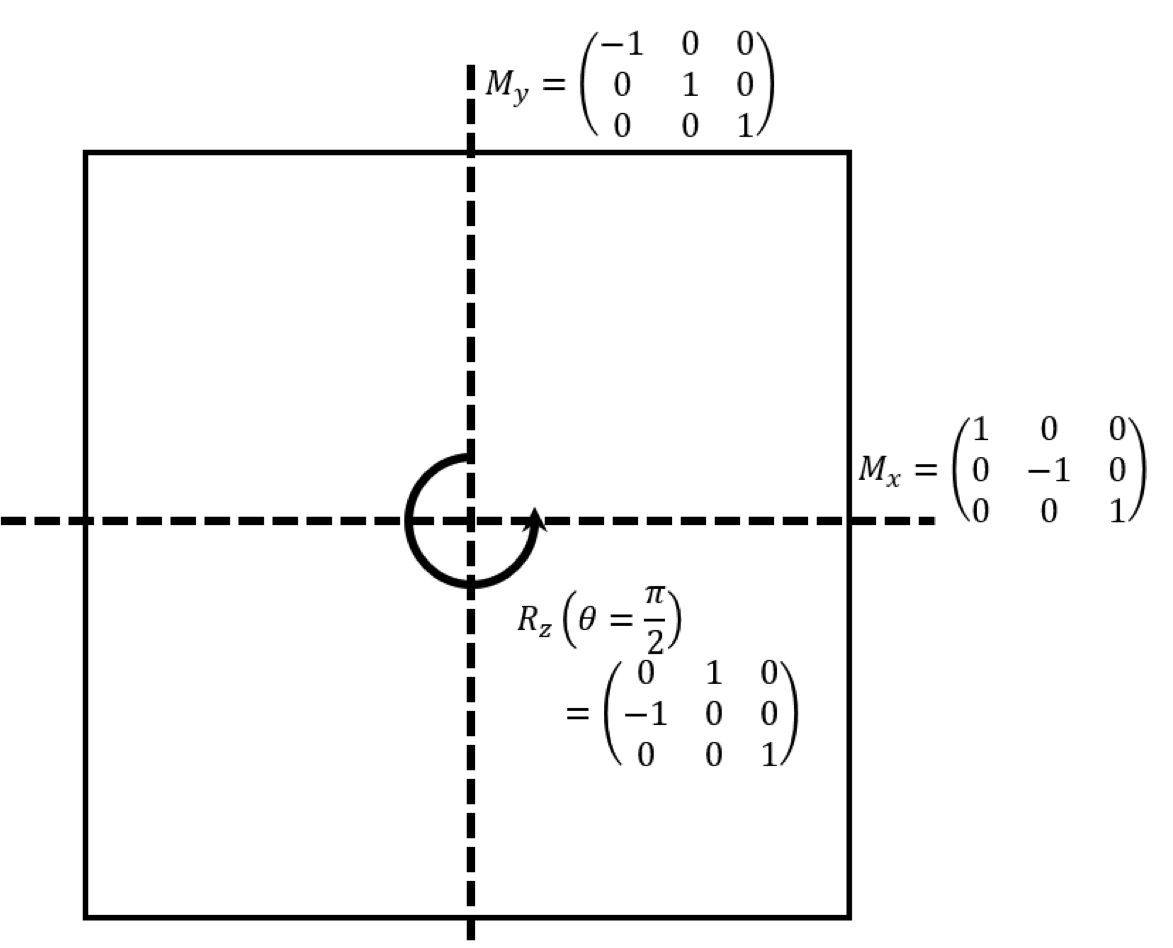}
    \caption{Symmetry operations of the $C_{4v}$ point group }
    \label{fig:C4v_Symm_Cartoon}
\end{figure}

Elucidating the underlying mechanism behind the helicity-dependent photocurrent in TaAs requires us to differentiate between the CPGE and CPDE by considering what is allowed by crystal symmetry. Given that TaAs belongs to the $C_{4v}$ point group, Neumann's principle dictates that the photocurrent response tensor be invariant under the symmetry operations of the crystal. This includes two mirror reflections, $M^{x}$ and $M^{y}$, and a $C_{4}=R(\frac{\pi}{2}\hat{z})$ rotation (Fig. \ref{fig:C4v_Symm_Cartoon}). In this case, mirror reflection about $\hat{x}$ yields, 
\begin{equation}
    \begin{split}
       J_{l}^{CPGE} & = M_{lk}^{x}J_{k}=iM_{lk}^{x}\gamma_{kp}(\vec{E}\times\vec{E^{*}})_{p}=iM_{lk}^{x}\gamma_{kp}M_{pn}^{x}M_{np}^{x}(\vec{E}\times\vec{E^{*}})_{p}\\&=-iM_{lk}^{x}\gamma_{kp}M_{pn}^{x}(\vec{E}\times\vec{E^{*}})_{n},
     \label{Mirror_X_CPGE} 
    \end{split}
\end{equation}
where the identity is inserted in the form of $M_{pn}^{x}M_{np}^{x}$ and a change in sign comes about due to acting on the pseudo-vector $(\vec{E}\times\vec{E}^{*})$ with an improper rotation. 

By relating equations (\ref{CPGE_Equation}) and (\ref{Mirror_X_CPGE}), the CPGE tensor transforms as   
\begin{equation}
     \gamma_{lk}=-M_{ln}^{x}\gamma_{nm}M_{mk}^{x},
     \label{Mirror_02}
\end{equation}
which when taken in concert with both a mirror reflection about $\hat{y}$, 
\begin{equation}
     \gamma_{lk}=-M_{ln}^{y}\gamma_{nm}M_{mk}^{y},
     \label{Mirror_Y_CPGE02}
\end{equation}
as well as a rotation about $\Hat{z}$ by $90^{\circ}$,
\begin{equation}
     \gamma_{lk}=R_{ln}(\frac{\pi}{2}\hat{z})\gamma_{nm}R_{mk}(-\frac{\pi}{2}\hat{z}),
     \label{Rot_CPGE}
\end{equation}
yields the following relation between the remaining non-zero elements of the CPGE tensor  
\begin{equation}
     \gamma_{xy}=-\gamma_{yx}.
     \label{CPGE_C4v}
\end{equation}
Similarly, under $C_{4v}$ symmetry, a relation between the remaining non-zero elements of the CPDE tensor will be given by 
\begin{equation}
    \begin{split}
     & T_{xyz} = -T_{yxz}\\
     & T_{xzy} = -T_{yzx},
    \end{split}
     \label{CPDE_C4v}
\end{equation}
where there is the added constraint that $q_{n}(\vec{E}\times\vec{E}^{*})_{s}=q_{s}(\vec{E}\times\vec{E}^{*})_{n}$ as $(\vec{E}\times\vec{E}^{*})$ is necessarily parallel to $q$ for a transverse electromagnetic wave.

For a conventional Cartesian basis, such as that defined on the (001) surface, transverse helicity-dependent photocurrents flowing within the $ab$-plane are allowed under the symmetry constraints imposed by the $C_{4v}$ point group on the CPGE tensor. In particular, after making use of equation (\ref{CPGE_C4v}) such photocurrents are given by,  
\begin{equation}
    \begin{split}
     & J_{x}^{CPGE}=i\gamma_{xy}(\vec{E}\times\vec{E}^{*})_{y}\\
     & J_{y}^{CPGE}=-i\gamma_{xy}(\vec{E}\times\vec{E}^{*})_{x}
     \label{Eqn:CPGE_001}
     \end{split}
\end{equation}
where light propagating parallel to either the [0,1,0] or [1,0,0] axes will generate photocurrents along the transverse [1,0,0], or [0,1,0] axes, respectively. In contrast, the generation of a \textit{helicity}-dependent photocurrent for light normally incident on the (001) surface is symmetry forbidden and will therefore yield a weak THz emission spectrum, as experimentally observed in the inset of Fig. 3(b) of the main text as well as Fig. \ref{fig:CPGE_001Face}(a) of the supplementary information. Taken together, the generation of a transverse helicity-dependent photocurrent within the $ab$-plane, as well as the fact that such a photocurrent is symmetry forbidden for light normally incident along the $c$-axis, leads to the empirical argument that $\hat{J}^{CPGE}\propto \hat{k}\times\hat{c}$ \cite{Ma2017_Supp}.

Similarly, transverse helicity-dependent photocurrents flowing within the $ab$-plane are allowed under the symmetry constraints imposed on the CPDE tensor. That is, by considering once again a Cartesian basis,
\begin{equation}
    \begin{split}
     & J_{x}^{CPDE}=i(T_{xyz}q_{y}(\vec{E}\times\vec{E}^{*})_{z}+T_{xzy}q_{z}(\vec{E}\times\vec{E}^{*})_{y})\\
     & J_{y}^{CPDE}=i(T_{yxz}q_{x}(\vec{E}\times\vec{E}^{*})_{z}+T_{yzx}q_{z}(\vec{E}\times\vec{E}^{*})_{x}),
     \label{Eqn:CPDE_001}
     \end{split}
\end{equation}
it follows that such photocurrents are generated so long as momentum transfer occurs along the direction of light propagation and must therefore lie within the scattering plane. In this way it is possible to readily distinguish between the CPDE and the CPGE by noting from equation (\ref{Eqn:CPDE_001}) that the CPDE is only allowed under an oblique angle of incidence. Thus, any helicity-dependent photocurrent generated at normal incidence must then be attributed to the CPGE. As this pertains to the (001) surface, however, the generation of transverse helicity-dependent photocurrents is symmetry forbidden for light propagating along the $c$-axis and, as shown in Fig. \ref{fig:CPGE_001Face}(b) of the supplementary information, only arises with obliquely incident light. Consequently, despite the fact that the empirical argument, $\hat{J}^{CPGE}\propto \hat{k}\times\hat{c}$, still holds in description of Fig. \ref{fig:CPGE_001Face}(b), it is not possible to unambiguously assign the origin of such a helicity-dependent photocurrent to either the CPGE or the CPDE.

In contrast to the (001) face, the (112) face allows for transverse helicity-dependent photocurrents generated by the CPGE to be clearly distinguished from those arising from the CPDE. To demonstrate this fact, we begin by transforming from the conventional Cartesian basis used above, $(\hat{x}, \hat{y}, \hat{z})$, to that which defines the high symmetry axes of the (112) surface \cite{Patankar2018}.
\begin{equation}
    \begin{split}
     & \lambda = \frac{1}{\sqrt{2}}(\hat{x}, -\hat{y}, 0),\\
     & \mu = \frac{1}{\sqrt{3}}(\hat{x}, \hat{y}, -\hat{z}),\\
     & \nu = \frac{1}{\sqrt{6}}(\hat{x}, \hat{y}, 2\hat{z}).
     \label{Eqn:Basis_112}
     \end{split}
\end{equation}
For light normally incident on the (112) face, it follows that $\vec{q}=q\hat{\nu}$, while $\vec{E}$ must lie in the $\lambda\mu$-plane. By inverting the transformation matrix defined in equation (\ref{Eqn:Basis_112}), both the $\vec{E}$-field and wavevector, $q$, can be expressed in terms of the independent quantities $E_{\lambda}$, $E_{\mu}$, 
\begin{equation}
    \begin{split}
     & E_{x} = \frac{1}{\sqrt{2}}E_{\lambda} + \frac{1}{\sqrt{3}}E_{\mu},\\
     & E_{y} = -\frac{1}{\sqrt{2}}E_{\lambda} + \frac{1}{\sqrt{3}}E_{\mu},\\
     & E_{z} = -\frac{1}{\sqrt{3}}E_{\mu},
     \label{Eqn:EField_112}
     \end{split}
\end{equation}
and $q_{\nu}$,
\begin{equation}
    \begin{split}
     & q_{x} = \frac{1}{\sqrt{6}}q_{\nu},\\
     & q_{y} = \frac{1}{\sqrt{6}}q_{\nu},\\
     & q_{z} = \sqrt{\frac{2}{3}}q_{\nu}.
     \label{Eqn:q_112}
     \end{split}
\end{equation}
As a result, the two orthogonal in-plane photocurrents, $J_{\lambda} = \frac{1}{\sqrt{2}}(J_{x}-J_{y})$ and $J_{\mu} = \frac{1}{\sqrt{3}}(J_{x}+J_{y}-J_{z})$, observed in this experiment can be written in terms of tunable quantities expressed in the $(\hat{\lambda},\hat{\mu},\hat{\nu})$ basis of the (112) surface. 

By substituting equation (\ref{Eqn:EField_112}) into equation (\ref{Eqn:CPGE_001}), expanding the circular photogalvanic tensor to the form of equation (\ref{CPGE_Equation}), and invoking equation (\ref{CPGE_C4v}), it can be shown that a  helicity-dependent photocurrent generated along the [1,-1,0] axis is allowed by symmetry of the CPGE tensor,
\begin{equation}
    \begin{split}
     J_{\lambda}^{CPGE} = \frac{1}{\sqrt{2}}(J_{x}-J_{y}) & = i\frac{\eta_{xzx}}{\sqrt{2}}(E_{z}E_{x}^{*}-E_{x}E_{z}^{*}+E_{y}E_{z}^{*}-E_{z}E_{y}^{*})\\
     & = i\frac{\eta_{xzx}}{\sqrt{3}}(E_{\lambda}E_{\mu}^{*}-E_{\mu}E_{\lambda}^{*}),\\
     & = i\frac{\gamma_{xy}}{\sqrt{3}}(\vec{E}\times\vec{E}^{*})_{\nu},
     \label{Eqn:CPGE_112_1n10}
     \end{split}
\end{equation}
while that flowing along the [1,1,-1] axis is symmetry forbidden, 
\begin{equation}
    \begin{split}
     J_{\mu}^{CPGE} = \frac{1}{\sqrt{3}}(J_{x}+J_{y}-J_{z}) & = i\frac{\eta_{xzx}}{\sqrt{3}}(E_{z}E_{x}^{*}-E_{x}E_{z}^{*}-E_{y}E_{z}^{*}+E_{z}E_{y}^{*})\\
     & = 0.
     \label{Eqn:CPGE_112_11n1}
     \end{split}
\end{equation}
In contrast, the opposite behavior is observed for the CPDE tensor, 
\begin{equation}
    \begin{split}
     & J_{\lambda}^{CPDE} = 0\\
     & J_{\mu}^{CPDE} = i \frac{2}{3\sqrt{3}}(T_{xyxy}+T_{xzxz})q_{\nu}(\vec{E}\times\vec{E}^{*})_{\nu},
     \label{Eqn:CPDE_112}
     \end{split}
\end{equation}
meaning that helicity-dependent photocurrents generated by either the CPGE or CPDE can be distinguished from one another based on the direction that current flows relative to the orthogonal high symmetry axes of the (112) surface. Here, experimental results shown in Fig. 1 and Fig. 2 of the main text clearly illustrate the generation of an ultrafast helicity-dependent photocurrent along [1,-1,0], while that generated along [1,1,-1] is found to be largely polarization independent. Thus, given that our experimental results are in excellent agreement what is expected by symmetry for the CPGE, we argue that the CPDE can be excluded as a possible origin for the ultrafast helicity-dependent photocurrent generated on the (112) face of TaAs. 

\bibliography{Sirica_Supp}